\documentstyle[12pt,epsfig]{article}                                
\begin{document}
\titlepage                  
\vspace{1.0in}
\begin{huge}
\begin{bf}
\begin{center}

 $\bf{PT}$-Symmetric Quantum Mechanics
  
\end{center}
\end{bf}
\end{huge}
\begin{center}
\begin{bf}
	      Biswanath Rath

\end{bf}

Department of Physics, North Orissa University, Takatpur, Baripada -757003, Odisha, INDIA(:E.mail:biswanathrath10@gmail.com).

\end{center} 
\vspace{0.5in}
We propose a model  $\bf{CCS}$(complex-conjugate-space)  to understand the
 inner and outer product nature of wave functions in  non-hermitian $\bf{PT}$-symmetry model in quantum mechanics considering $(N\times N)$ matrix model .
Further we reflect the correct nature of C-symmetry ,P-parity , T-time reversal in matrix form and  the orignal Hamiltonian matrix for any arbitrary values of N. Interestingly the present result on C and P for  N=2 , remains the same reported earlier by Bender,Brody and Jones model $\bf{PT}$-symmetry operaor . In non-conventional way one can notice that wave functions in a $\bf{PT}$-symmetry model satifies similar relations as in hermitian operator .

\vspace{0.1cm}

PACS no-03.65.Db

\vspace{0.1cm}

Key words- $\bf{PT}$-symmetry ,Hermiticity ,wave functions , properties ,C-symmetry 

\begin{bf}
\hspace{2.0in}  1.Introduction
\end{bf}

Since the development of quantum mechanics it is well known that Wave functions in Hermitian operator satisfies the following relations[1,2]
\begin{equation}
<\psi_{i}||\psi_{j}>=\delta_{i,j}
\end{equation}
\begin{equation}
<\psi_{i}|H=H^{\dagger}|\psi_{i}>=E_{i}
\end{equation}
\begin{equation}
H=\sum_{i}E_{i}|\psi_{i}><\psi_{i}|
\end{equation}
\begin{equation}
I=\sum_{i}|\psi_{i}><\psi_{i}|
\end{equation}
In above $E_{i}$ is real and $I$ is an unit matrix having dimension (NxN).
Interestingly understanding of inner oand outer product as reflected above 
is very easy . However if $\bf{Bra}:<\psi|$ [1,3-8] is  different from $\bf{ket}$ ($|>$)
then above analysis can hardly be valid . Quantum mechanical model where $\bf{Bra}$ is different from $\bf{ket}$ are seen(in the $\bf{PT}$-symmetric model of 
Hamiltonian satisfying the condition 
\begin{equation}
[H,PT]=0 \rightarrow H\neq H^{\dagger}
\end{equation}
.

 It should be borne in mind that all the Hamiltonian satisfying the above condition not necessarily give real spectra . Hence $\bf{PT}$-symmetry model can be 
categorized as

 (i) un-broken $\bf{PT}$-symmetry and (ii) broken $\bf{PT}$-symmetry . Here

$P$ stands for parity operator having the behaviour

\begin{equation}
P(x) \rightarrow -x 
\end{equation}
\begin{equation}
P(i) \rightarrow i
\end{equation}
 Similarly $\bf{T}$-stands for time -reversal operator having the behaviour[2]
\begin{equation}
T(i) \rightarrow -i
\end{equation}
and 
\begin{equation}
T(x) \rightarrow x
\end{equation}
In other words $\bf{PT}$ stands for parity-time reversal symmetry [3] .
It interesting to discuss all previous models in the derivation of appropriate 
inner product and outer product model authors [3,4]have argued that 
\begin{equation}
<\psi|  = <\psi^{CPT}|
\end{equation}
Further some authors[6]
 have argued that CPT-product means 
\begin{equation}
<\psi^{*}|\psi>=\int \psi^{2} dx  
\end{equation}
In above $C$ is the C-symmetry operator having eigenvalues $\lambda_{C}=\pm 1$  satisfying the condition
\begin{equation}
[H,C]=0
\end{equation}
and 
$P$-is the parity operator having eigenvalues $\lambda_{P}=\pm 1$ . Both of these operators are determined from the original Hamiltonian operator [4] .For example consider the Hamiltonian 
\begin{equation}
H=\sigma_{0} r \cos \theta + \sigma \dot (s,0,s \sin \alpha)
\end{equation}
where $\sigma_{1,2,3}$ are the Pauli matrices and $\sigma_{0}$=unit matrix. The Parity operator is 
\begin{equation}
P=\sigma_{1}  
\end{equation}
and 
\begin{equation}
C=  \sigma \dot (s,0,s \sin \alpha) (s=\sec \alpha)
\end{equation}
Here the outer product is defined as 
\begin{equation}
<\psi^{CPT}_{\pm}| | \psi_{\pm}=1 ; <\psi^{CPT}_{\pm}| | \psi_{\mp}>=0
\end{equation}
On the other hand in Wang model [5] ,one defines the outer product as 
\begin{equation}
<\psi_{\pm}|W|\psi_{\pm}>=1 ; <\psi_{\pm}|W|\psi_{\mp}>= 0
\end{equation}
where $W=(R^{-1})^{\dagger} W_{0} R^{-1}$ ( for details see the reference[5].
In contract to these two model , Manneheim[7] discussed the outer product as 
\begin{equation}
<\psi_{\pm}|V|\psi_{\pm}>=1 ; <\psi_{\pm}|V|\psi_{\mp}>= 0
\end{equation}
In comparing above models we notice that all the approaches are different from each other  in describing the same physics . However all the previous models are discussed considering $(2\times 2)$ matrix model and no discussion has been given for $(N\times N)$ model of non-Hermitian model . On the other hand 
we feel that inner and outer product in complex $\bf{PT}$-symmetry can be described eleglantly without the help of previous assumptions as follows .
The presentation is as follows : first we discuss the model considering a 
 $(2\dot 2)$ matrix , then extend it for higher values of N say N=3,4,5,10 .
The proposed matrix model considered here is[1]
\begin{equation}
 H^{PT} = \left[
\begin{array}{cc}
re^{i\theta} & s  \\
s & re^{-i\theta} \\
\end{array}
\right ]
\end{equation}
It is actually a three parameter model. The  model under unbroken eigenvalues[2] are the following 
\begin{equation}
E_{1,2}= r \cos \theta \pm s \cos \phi
\end{equation}
and the corresponding eigenfunctions are [6]
\begin{equation}
|\psi> = \frac{1}{\sqrt{2\cos \phi}}     \left[
\begin{array}{c}
e^{i\phi/2}  \\
e^{-i\phi/2} \\
\end{array}
\right ]
\end{equation}
and
\begin{equation}
|\phi> = \frac{1}{\sqrt{2\cos \phi}}     \left[
\begin{array}{c}
e^{-i\phi/2}  \\
-e^{i\phi/2} \\
\end{array}
\right ]
\end{equation}
In above relations one has to use $r\sin \theta = s \sin \phi$.
It is easy to see that in normal understanding of quantum mechanics ,the above wave functions are not normalised so 
$<\psi|H|\psi> \neq  E_{1}$ or $ <\phi|H|\phi> \neq E_{2} $.
Now we will consider a space, where  the above relations will remain valid similar to to one in standard Hilbert's space understanding.

\begin{bf}
\hspace{2.0in} 2.Complex Conjugate Space (CCS)
\end{bf}

 In order to accept the above wave functions pertaining to 
$\bf{PT}$-symmetry ,let us adopt the complex-conjugate -space($\bf{CCS}$) .In complex-conjugate-space ($\bf{CCS}$) the normalisation condition is  

\begin{bf}
Normalisation condition
\end{bf}

\begin{equation}
<\psi^{*}||\psi>=<\phi^{*}||\phi>=1
\end{equation}
Interested readers can verify that 
\begin{equation}
<\psi^{*}|=\frac{1}{\sqrt{2\cos \phi  }}[e^{i\phi /2},e^{-i\phi/2}]
\end{equation}
\begin{equation}
<\phi^{*}|=\frac{1}{\sqrt{2\cos \phi  }}[e^{-i\phi /2},-e^{i\phi/2}]
\end{equation}
and the corresponding normalisation condition as proposed above . Under the 
complex-conjugate-space  the other lations are the following 

\begin{bf}
Orthogonality condition
\end{bf}

\begin{equation}
<\psi^{*}||\phi>=0=<\phi^{*}||\psi>
\end{equation}

\begin{bf}
Energy calculation
\end{bf}

\begin{equation}
<\psi^{*}|H^{PT}|\psi>=E_{1}=r \cos \theta + s \cos \phi
\end{equation}
\begin{equation}
<\phi^{*}|H^{PT}|\phi>=E_{2}=r \cos \theta - s \cos \phi
\end{equation}

\begin{bf}
Original Matrix
\end{bf}

Then it is easy to see that 

\begin{equation}
H^{PT}=E_{1}|\psi> <\psi^{*}| + E_{2}|\phi><\phi^{*}|= \left[
\begin{array}{cc}
re^{i\theta} & s  \\
s & re^{-i\theta} \\
\end{array}
\right ]
\end{equation}

\begin{bf}
Identity operator (Completeness relation )
\end{bf}

\begin{equation}
I=|\psi> <\psi^{*}| + |\phi><\phi^{*}| = \left[
\begin{array}{cc}
1 & 0  \\
0 & 1 \\
\end{array}
\right ]
\end{equation}

\begin{bf}
C-Symmetry operator
\end{bf}

 In hermiticity we do not give importance to another symmetry called C-symmetry having eigenvalues $\pm 1$ . In $\bf{PT}$-symmetry one can calculate 
$C$ as 
\begin{equation}
C=|\psi> <\psi^{*}|-|\phi><\phi^{*}| = \left[
\begin{array}{cc}
i \tan \phi & \sec \phi  \\
\sec \phi & -i\tan \phi \\
\end{array}
\right ]
\end{equation}

Interested reader can verify that 
\begin{equation}
<\psi^{*}|C|\psi>=\lambda_{1}=1
\end{equation}
\begin{equation}
<\phi^{*}|C|\phi>=\lambda_{2}=-1
\end{equation}

\begin{bf}
P-Parity operator 
\end{bf}

In this context, the  parity operator can be written as [5] 
\begin{equation}
P= \sum_{i}|\psi_{i}><\psi_{i}|^{-1} C = 
\left[
\begin{array}{cc}
0 & 1  \\
1 & 0 \\
\end{array}
\right ]
\end{equation}
or 
\begin{equation}
P=|\psi^{*}> <\psi^{*}|-|\psi^{*}><\psi^{*}| =
\left[
\begin{array}{cc}
0 & 1  \\
1 & 0 \\
\end{array}
\right ]
\end{equation}

\begin{bf}
T -Time- reversal operator 
\end{bf}

In fact T-time reversal operator can be determined from the PT-symmetry telation as 
\begin{equation}
 [H,PT]=0 \Longleftrightarrow   T^{-1} (P^{-1} H P) T = H
\end{equation}
considering [1,2]
\begin{equation}
T =  A K
\end{equation}
where A is a  (2x2) matrix 

\begin{equation}
A =
\left[
\begin{array}{cc}
a & b  \\
d & d \\
\end{array}
\right ]
\end{equation}

and 
K is a complex conjugation operator with 
\begin{equation}
K^{-1}=K , 
\end{equation}
Considering above T can be expressed as 
 
\begin{equation}
T=
\left[
\begin{array}{cc}
1 & 0  \\
0 & 1 \\
\end{array}
\right ]K
\end{equation}

Hence it is easy to prove the CPT relation as 
\begin{equation}
 [H,CPT]=0 \Longleftrightarrow   T^{-1} [(P^{-1} [C^{-1}H C] P ]T = H
\end{equation}

\begin{bf}
\hspace{2.0in} 3.Higher values on N
\end{bf}

Here we consider the operator $C$ and $P$ for higher order matrix having 
dimension $N\gg 2$ The operator P can be expressed as 
\begin{equation}
P=\sum_{n}(-1)^{n}|\psi^{*}_{n}><\psi^{*}_{n}|
\end{equation}
and $C$-symmetry operator can be expressed as 
\begin{equation}
C=\sum_{n} (-1)^{n}|\psi_{n}><\psi^{*}_{n}|=\sum_{n}(-1)^{n}\psi_{n} \otimes \psi^{\dagger}_{n}
\end{equation}
\begin{equation}
|\psi_{+}><\psi^{*}_{+}|-|\psi_{-}><\psi^{*}_{-}|=C
\end{equation}
and indentity operator as 
\begin{equation}
|\psi^{*}_{+}><\psi_{+}|+|\psi^{*}_{-}><\psi_{-}|= I
\end{equation}
Now apply it to for different size of matrix as follows .

\begin{bf}
N=3
\end{bf}

Let us apply to $H_{3}^{PT}$ 
\begin{equation}
 H_{3}^{PT} = \left[
\begin{array}{ccc}
r_{1}e^{i\theta_{1}} & s_{1} & 0  \\
s_{1} & r_{1}e^{-i\theta_{1}}& 0 \\
0&0&a \\
\end{array}
\right ]
\end{equation}

with known solutions as 

\begin{equation}
|\psi^{(1)}_{+}> = \frac{1}{\sqrt{2\cos \phi_{1}}}     \left[
\begin{array}{c}
e^{i\phi_{1}/2}  \\
e^{-i\phi_{1}/2} \\
0 \\
\end{array}
\right ]
\end{equation}

\begin{equation}
|\psi^{(1)}_{-}> = \frac{1}{\sqrt{2\cos \phi_{1}}}     \left[
\begin{array}{c}
e^{-i\phi_{1}/2}  \\
-e^{i\phi_{1}/2} \\
0  \\
\end{array}
\right ]
\end{equation}

\begin{equation}
|\psi_{a}> =     \left[
\begin{array}{c}
0  \\
0 \\
1  \\
\end{array}
\right ]
\end{equation}

where $r_{1} \sin \theta_{1} = s_{1} \sin \phi_{1}$ ( the condition for unbroken spectra . Then it is easy to see that 
\begin{equation}
<\psi^{(1)*}_{+}|H_{1}^{PT}|\psi^{(1)}_{+}>=E^{(1)}_{+}=r_{1} \cos \theta_{1} + s_{1} \cos \phi_{1}
\end{equation}
\begin{equation}
<\psi^{(1)*}_{-}|H_{1}^{PT}|\psi^{(1)}_{-}>=E^{(1)}_{-}=r_{1} \cos \theta_{1} + s_{1} \cos \psi_{1}
\end{equation}
\begin{equation}
<\psi_{a}|H_{1}^{PT}|\psi_{a}>= a
\end{equation}

\begin{equation}
C_{3}=
|\psi^{(1)}_{+}> <\psi^{(1)*}_{+}|-|\psi^{(1)}_{-}><\psi^{(1)*}_{-}|+|\psi_{a}><\psi_{a}| 
= \left[
\begin{array}{ccc}
i \tan \phi_{1} & \sec \phi_{1}& 0  \\
\sec \phi_{1} & -i\tan \phi_{1}&0 \\
0&0&1 \\
\end{array}
\right ]
\end{equation}
\begin{equation}
P_{3}=|\psi^{(1)*}_{+}> <\psi^{(1)*}_{+}| - |\phi^{(1)*}_{-}><\phi^{(1)*}_{-}|+ |\psi_{a}><\psi_{a}| =
\left[
\begin{array}{ccc}
0 & 1 & 0 \\
1 & 0 & 0 \\
0 &0 &1 \\
\end{array}
\right ]
\end{equation}

\begin{equation}
T_{3}=
\left[
\begin{array}{ccc}
1 & 0 & 0 \\
0 & 1 & 0 \\
0 &0 &1 \\
\end{array}
\right ]K
\end{equation}

\begin{bf}
N=4
\end{bf}

Let us apply to $H_{4}^{PT}$ 
\begin{equation}
 H_{4}^{PT} = \left[
\begin{array}{cccc}
r_{1}e^{i\theta_{1}} & s_{1} & 0 & 0 \\
s_{1} & r_{1}e^{-i\theta_{1}}& 0 &0 \\
0&0&r_{2}e^{i\theta_{2}} & s_{2}   \\
0&0&s_{2} & r_{2}e^{-i\theta_{2}}\\
\end{array}
\right ]
\end{equation}

having solutions 

\begin{equation}
|\psi^{(1)}_{+}> = \frac{1}{\sqrt{2\cos \phi_{1}}}     \left[
\begin{array}{c}
e^{i\phi_{1}/2}  \\
e^{-i\phi_{1}/2} \\
0\\
0\\
\end{array}
\right ]
\end{equation}

\begin{equation}
|\psi^{(1)}_{-}> = \frac{1}{\sqrt{2\cos \phi_{1}}}     \left[
\begin{array}{c}
e^{-i\phi_{1}/2}  \\
-e^{i\phi_{1}/2} \\
0  \\
0 \\
\end{array}
\right ]
\end{equation}

\begin{equation}
|\psi^{(2)}_{+}> = \frac{1}{\sqrt{2\cos \phi_{2}}}     \left[
\begin{array}{c}
0\\
0\\
e^{i\phi_{2}/2}  \\
e^{-i\phi_{2}/2} \\
\end{array}
\right ]
\end{equation}
and
\begin{equation}
|\psi^{(2)}_{-}> = \frac{1}{\sqrt{2\cos \phi_{2}}}     \left[
\begin{array}{c}
0\\
0\\
e^{-i\phi_{2}/2}  \\
-e^{i\phi_{2}/2} \\
\end{array}
\right ]
\end{equation}
where $r_{i} \sin \theta_{i} = s_{i} \sin \phi_{i}$ (i=1,2 the condition for unbroken spectra . Then it is easy to see that 
\begin{equation}
<\psi_{\pm}^{(i)}|H^{PT}_{4}|\psi_{\pm}^{(i)}>=E^{(i)}_{\pm}=r_{i}\cos \theta_{i} \pm s_{i} \cos \phi_{i}
\end{equation}
\begin{equation}
C_{4}=\sum_{i}[|\psi_{+}^{(i)}><\psi_{+}^{(i*)}|-|\psi_{-}^{(i)}><\psi_{-}^{(i*)}| ]= \left[
\begin{array}{cccc}
i \tan \phi_{1} & \sec \phi_{1}& 0 &0 \\
\sec \phi_{1} & -i\tan \phi_{1}&0&0 \\
0&0&i \tan \phi_{2} & \sec \phi_{2} \\
0&0&\sec \phi_{2} & -i\tan \phi_{2} \\
\end{array}
\right ]
\end{equation}
\begin{equation}
P_{4}=\sum_{i}[|\psi_{+}^{(i*)}><\psi_{+}^{(i*)}|-|\psi_{-}^{(i*)}><\psi_{-}^{(i*)}| ]=
\left[
\begin{array}{cccc}
0 & 1 & 0 &0\\
1 & 0 & 0 &0\\
0 & 0 & 0 &1\\
0 & 0 & 1 &0\\
\end{array}
\right ]
\end{equation}
 
\begin{equation}
T_{4}=
\left[
\begin{array}{cccc}
1 & 0 & 0 &0\\
0 & 1 & 0 &0\\
0 & 0 & 1 &0\\
0 & 0 & 0 &1\\
\end{array}
\right ]K
\end{equation}

\begin{bf}
N=5
\end{bf}

Similarly for 
\begin{equation}
 H_{5}^{PT} = \left[
\begin{array}{ccccc}
r_{1}e^{i\theta_{1}} & s_{1} & 0 & 0&0 \\
s_{1} & r_{1}e^{-i\theta_{1}}& 0&0&0 \\
0&0&r_{2}e^{i\theta_{2}} & s_{2} & 0  \\
0&0&s_{2} & r_{2}e^{-i\theta_{2}} & 0\\
0&0&0&0&a \\
\end{array}
\right ]
\end{equation}
one can have  
\begin{equation}
C_{5}= \left[
\begin{array}{ccccc}
i \tan \phi_{1} & \sec \phi_{1}& 0 &0 & 0 \\
\sec \phi_{1} & -i\tan \phi_{1}&0&0 & 0\\
0&0&i \tan \phi_{2} & \sec \phi_{2}& 0 \\
0&0&\sec \phi_{2} & -i\tan \phi_{2}& 0  \\
0&0&0&0& 1 \\
\end{array}
\right ]
\end{equation}
\begin{equation}
P_{5}=
\left[
\begin{array}{ccccc}
0 & 1 & 0 &0 & 0\\
1 & 0 & 0 &0 & 0\\
0 & 0 & 0 &1&0 \\
0 & 0 & 1 &0&0\\
0&0&0&0& 1 \\
\end{array}
\right ]
\end{equation}
\begin{equation}
T_{5}=
\left[
\begin{array}{ccccc}
1 & 0 & 0 &0 & 0\\
0 & 1 & 0 &0 & 0\\
0 & 0 & 1 &0&0 \\
0 & 0 & 0 &1&0\\
0&0&0&0& 1 \\
\end{array}
\right ] K
\end{equation}

\begin{bf}
N=10
\end{bf}

Let us define $\tau_{j} = r_{j}e^{i\theta_{j}}$;$\sigma_{j} = r_{j}e^{-i\theta_{j}}$; $ r_{j}\sin \theta_{j} = s_{j} \sin \phi_{j}$.

\begin{equation}
 H_{10}^{PT} = \left[
\begin{array}{cccccccccc}
\tau_{1} & s_{1} & 0 & 0&0 &0 &0 &0 &0 &0  \\
s_{1} & \sigma_{1}& 0&0&0&0&0&0&0& 0\\
0& 0& \tau_{2} & s_{2} & 0 & 0&0 &0 &0 & 0  \\
0& 0& s_{2} & \sigma_{2}& 0&0&0&0 & 0&0 \\
0&0&0&0&\tau_{3} & s_{3} & 0 & 0&0 & 0  \\
0& 0& 0& 0& s_{3} & \sigma_{3}& 0&0&0&0 \\
0&0&0&0&0&0&\tau_{4} & s_{4} & 0 & 0 \\
0&0&0&0&0&0& s_{4} & \sigma_{4}& 0 & 0  \\
0&0&0&0&0&0&0&0&\tau_{5} & s_{5} \\
0&0&0&0&0&0&0&0&s_{5} & \sigma_{5} \\
\end{array}
\right ]
\end{equation}
 In order to present the expression for $\bf{C_{10}}$ we define :$\Omega_{j}=i\tan \phi_{j};\Lambda_{j}=\sec \phi_{j}$

\begin{equation}
 C_{10} = \left[
\begin{array}{cccccccccc}
\Omega_{1} & \Lambda_{1} & 0 & 0&0 &0 &0 &0 &0 &0  \\
\Lambda_{1} & -\Omega_{1}& 0&0&0&0&0&0&0& 0\\
0& 0& \Omega_{2} & \Lambda_{2} & 0 & 0&0 &0 &0 & 0  \\
0& 0& \Lambda_{2} & -\Omega_{2}& 0&0&0&0 & 0&0 \\
0&0&0&0&\Omega_{3} & \Lambda_{3} & 0 & 0&0 & 0  \\
0& 0& 0& 0& \Lambda_{3} & -\Omega_{3}& 0&0&0&0 \\
0&0&0&0&0&0&\Omega_{4} & \Lambda_{4} & 0 & 0 \\
0&0&0&0&0&0& \Lambda_{4} & -\Omega_{4}& 0 & 0  \\
0&0&0&0&0&0&0&0&\Omega_{5} & \Lambda_{5} \\
0&0&0&0&0&0&0&0&\Lambda_{5} & -\Omega_{5} \\
\end{array}
\right ]
\end{equation}

\begin{equation}
 P_{10} = \left[
\begin{array}{cccccccccc}
0 & 1 &  & & & & & & &  \\
1 & 0 &  & & & & & & & \\
& &0 & 1&  & & & & &   \\
& & 1 &0 & &&& & & \\
&&&&0 & 1 &  & & &   \\
& & & & 1 & 0& &&& \\
&&&&&&0 & 1 &  &  \\
&&&&&& 1&0 &  &   \\
&&&&&&&&0 &1 \\
&&&&&&&&1 & 0 \\
\end{array}
\right ]
\end{equation}

\begin{equation}
 T_{10} = \left[
\begin{array}{cccccccccc}
1 & 0 &  & & & & & & &  \\
0 & 1 &  & & & & & & & \\
& &1 & 0&  & & & & &   \\
& & 0 &1 & &&& & & \\
&&&&0 & 1 &  & & &   \\
& & & & 0 & 1& &&& \\
&&&&&&1 & 0 &  &  \\
&&&&&& 0&1 &  &   \\
&&&&&&&&1 &0 \\
&&&&&&&&0 & 1 \\
\end{array}
\right ]K
\end{equation}

\begin{bf}
\hspace{2.0in} 4.Conclusion
\end{bf}

In this paper we show that in general $C$, $P$ and $T$ can be calculated from the
 known wave functions using appropriate exression (for any matrix size (NxN);N=2,3,4 ,...,,) . Further both C and P have
 the same eigenvalues $\pm 1$. In general both are not trace less operators .
Interestingly P is always a real operator where as ,C may be complex or real-complex depending upon the nature of matrix size.C can also be calculated 
following the literature[1].We also believe a simple procedure is 
required in view of its interesting applications [8]. Lastly present 
symmetric and asymmetric case will improve the understanding of non-hermitian operators in quantum mechanics . Following the above procedure one can calculate T for any size of N. It is interesting to note that in any dimension (NxN) PT,CPT do not possess any eigenvalues as the said operator in combined form can not be expressed as a matrix (due to antilinear property ) . The only operator C has eigenvalues as it is expressed in matrix . Further a new symmetry can be 
generated as follows(1]  
\begin{equation}
HC=CH \Longleftrightarrow C^{-1}HC = H \Longleftrightarrow  [H,C]=0
\end{equation}

\begin{equation}
HC=CH \Longleftrightarrow  CHC^{-1}=H  \Longleftrightarrow HC^{-1}=HC^{-1}\Longleftrightarrow  [H,C^{-1}]=0
\end{equation}
Similarly it is not difficult to show 
\begin{equation}
[H,\frac{C}{\beta+C}]=0
\end{equation}

\begin{equation}
[H,\frac{C}{\beta+\frac{C}{\beta+C} }]=0
\end{equation}
In  general 
\begin{equation}
[H,F]
\end{equation}
where[7] 
\begin{equation}
F=\frac{C}{\beta + \frac{C}{\beta +  \frac{C}{\beta +\frac{ C }{\beta + \frac{C}{\beta + \frac{C}{\beta + \frac{C}{ \beta + \frac{C}{\beta + \frac{C}{\beta +\frac{ C }{ \beta + C}                  } }  } } } } } } }
\end{equation}

\begin{bf}
Declaration
\end{bf}

Present paper is a minor  modified form of reported work in arXiv by the author[1].

\vspace{1.0cm}
Journal Ref-The African Review of Physics(2019) $\bf{14}:0019$,139-143.
\end{document}